\documentclass[12pt,psfig]{article}
\usepackage{amsmath}
\usepackage{amsfonts}
\usepackage{graphics}
\usepackage{graphicx}
\usepackage{latexsym}
\usepackage{amsthm}
\usepackage{cite}
\usepackage{amssymb}
\usecounter{enumi}
\newtheorem{Theorem}{Theorem}
\newtheorem{Remark}{Remark}
\newtheorem{Lemma}{Lemma}
\newtheorem{Definition}{Definition}

\newtheorem*{Proof}{Proof}

\numberwithin{equation}{section}

\def\be{\begin{eqnarray}} \def\ee{\end{eqnarray}} 
  \def\({\left(} \def\){\right)}
\def\bc{\begin{center}} 
\def\ec{\end{center}}  

\def\bey{\begin{eqnarray*}}\def\eey{\end{eqnarray*}}
\begin{document}
\title{Pseudo-potentials and local isometric immersion for a generalized Camassa-Holm equation describing pseudospherical surfaces}

\author{ Mingyue Guo$^{a}$, Zhenhua Shi$^{a,b}$ \footnote{Corresponding author, E-mail address:
zhenhuashi@nwu.edu.cn}
\vspace{4mm}\\
$^{a}$\small School of Mathematics, Northwest University, Xi'an 710069, China\\
$^{b}$\small Center for Nonlinear Studies, Northwest University, Xi'an 710069, China}
\date{}
\maketitle
 \bc
\begin{minipage}{120mm}
{\bf Abstract}\\
A generalized Camassa-Holm equation, which describes pseudospherical surfaces, is considered. Using geometric methods, it is demonstrated that the equation is geometrically integrable. Additionally, an infinite hierarchy of conservation laws is derived. Furthermore, the paper delves into the investigation of the problem of locally isometric immersion into three-dimensional Euclidean space based on the equation.

{\bf Keywords:} equation describing pseudo-spherical surfaces, conservation laws, geometrical integrability, local isometric immersion
\end{minipage}
\ec

\section{Introduction}

\indent \indent Recently, Novikov in \cite{novikov2009generalizations} applied the perturbative symmetry approach to isolate and classify a broad class of integrable non-evolutionary partial differential equations of the form
\begin{equation}\label{1.1}
(1-\partial_x^2)u_t=F(u,u_x,u_{xx},u_{xxx}),
\end{equation}
where $u=u(x,t)$ and $F$ is some function of $u$ and its derivatives with respect to $x$. The classification results in 28 equations, including well-known ones such as the Camassa-Holm equation \cite{camassa1993integrable}, Degasperis-Procesi equation \cite{degasperis1999asymptotic},  Novikov equation \cite{hone2008integrable}. And some of these equations seem to be new and have yet to receive significant attention in research. Novikov also proposed the following integrable quasi-linear scaler evolution equation
\begin{equation}\label{1.2}
(1-\epsilon^2\partial_x^2)u_t=(1+\epsilon \partial_x)(\epsilon u^2u_{xx}+\epsilon uu_x^2-2u^2u_x),
\end{equation}
where $\epsilon\neq0$ is a real constant. It was shown in \cite{novikov2009generalizations} that Equation (1.2) possesses a hierarchy of local higher symmetries. By performing the transformation $(t,x) \mapsto (\epsilon t,\epsilon x)$,  one can obtain an equivalent form of Equation (1.2). In this paper, we focus on studying its equivalent form
\begin{equation}\label{1.3}
 u_t - u_{xxt} = u^2 u_{xxx} - u^2u_{xx}-3uu_x^2-2u^2u_x+4u u_x u_{xx}+u_x^3
\end{equation}
which describe pseudospherical surfaces with associated one-forms
\begin{equation}\label{1.4}
  \omega_1 = f_{11}dx + f_{12}dt, \quad \omega_2 = f_{21}dx + f_{22}dt, \quad \omega_3 = f_{31}dx + f_{32}dt,
\end{equation}
under the assumption, similar to that presented in \cite{silva2015third}, that
\begin{equation}\label{1.5}
 f_{p1} = \mu_p f_{11} + \eta_p, \quad \mu_p, \eta_p \in \mathbb{R}, \quad p = 2, 3.
\end{equation}

The geometric approach introduced by  S. S. Chern and K. Tenenblat \cite{chern1986pseudospherical} in 1986 and recently pursed by the author \cite{reyes2002geometric, reyes2005nonlocal, reyes2006pseudo, ablowitz1974inverse,reyes2000some}, is profitably applied to the study of Equation (1.3).  The result of our previous work demonstrates that Equation (1.3) describes pseudospherical surfaces as a special case. Consequently, it is straightforward to derive sequences of conservation laws from this observation. Furthermore, this paper also shows that Equation (1.3) possesses the property of admitting local isometric immersions in $\mathbb{E}^3$ with the "universal" coefficients of the second fundamental form, which depend solely on $x$ and $t$.

The organization of the paper can be summarized as follows: Section 2, we collect some results from references to complete the scheme of geometric integrability of Equation (1.3) by showing that it describes pseudospherical surfaces. Next, in Section 3, we recall the concept of a quadratic pseudo-potential and use this concept to derive a series of conservation laws for Equation  (1.3). In Section 4, we investigate the local isometric immersions of pseudospherical surfaces generated by the solutions of Equation (1.3).

\section{Geometric results}

\indent \indent For the reader's convenience, we briefly review some fundamental facts from the theory of equations that describe pseudospherical surfaces, which will be referenced throughout the paper.

If $ (M, g) $ is a 2-dimensional Riemannian manifold and $ \omega_1, \omega_2 $ is a coframe, dual to an orthogonal frame $ e_1, e_2 $, then the metric is given by $ g = \omega_1^2 + \omega_2^2 $ with $ \omega_i $ satisfy the structure equations: $ d\omega_1 = \omega_3 \wedge \omega_2 $ and $ d\omega_2 = \omega_1 \wedge \omega_3 $, where $ \omega_3 $ denotes the connection form defined as $ \omega_3(e_i) = d\omega_i(e_1, e_2) .$ The Gaussian curvature of $ M $ is the function $ K $ such that $ d\omega_3 = -K\omega_1 \wedge \omega_2 .$

 \begin{Definition}
 A $ k $-th order differential equation, for a real-valued function $u(x,t)$, describes pseudospherical surfaces, if it corresponds to the structure equations of a surface with Gaussian curvature $ K = -1 ,$ is given by:
\begin{equation}
	\begin{aligned}
		d\omega_1 &= \omega_3 \wedge \omega_2, \\
		d\omega_2 &= \omega_1 \wedge \omega_3, \\
		d\omega_3 &= \omega_1 \wedge \omega_2,
	\end{aligned}
\end{equation}
where $ \omega_1, \omega_2, \omega_3 $ are 1-forms $ \omega_i = f_{i1}dx + f_{i2}dt ,$ $ 1 \leq i \leq 3 ,$ with $ \omega_1 \wedge \omega_2 \neq 0 $ and $ f_{ij}, j = 1, 2 ,$ are functions of $ u(x,t) $ and its derivatives with respect to $x$ and $t$.
\end{Definition}

Notice that, by definition, the wedge product $ \omega_1 \wedge \omega_2 $ is non-vanishing for solutions $ u(x, t) $ to equations describing pseudospherical surfaces. However, this condition alone does not guarantee the desired property for all solutions $ u: U \subseteq \mathbb{R}^2 \rightarrow \mathbb{R} $, which is that $ \omega_1 \wedge \omega_2 $ is non-zero everywhere on $ U $. When referring to a system of 1-forms $ \omega_1, \omega_2, \omega_3 $, we define a generic solution as one for which $ \omega_1 \wedge \omega_2 $ is almost everywhere non-zero on $ U $, that is, it is non-zero except possibly on a set of measure zero. For any generic solution $ u(x,t) $ of an equation describing pseudospherical surfaces, the metric $ g = \omega_1^2 + \omega_2^2 $ defined by $ u $ describes a Riemannian metric on $ U $ with Gaussian curvature $ K = -1$ almost everywhere. This intrinsic geometric property of a Riemannian metric is significant and remains independent of the embedding in an ambient space.

 \begin{Definition}
 An equation is geometrically integrable if it describes a non-trivial one parameter family of pseudospherical surfaces.
  \end{Definition}

We begin our analysis by citing a result proved in our other work.
\begin{Lemma}
An equation of type
\begin{equation}
  u_t-u_{xxt}=\lambda u^2 u_{xxx}+G(u,u_x,u_{xx}),
\end{equation}
describes pseudospherical surfaces, with corresponding functions $f_{ij}=f_{ij}(u,u_x,...,\partial^k_x u)$ satisfying auxiliary conditions (1.5) with $(\mu_{3}\phi_{12}-\phi_{32})-\mu_{2}(\mu_{2}\phi_{32}-\mu_{3}\phi_{22})=0$ and $\phi_{22}-\mu_{2}\phi_{12}\neq0$, if and only if it can be written in the following form
\begin{equation*}
\begin{split}
&G=\frac{1}{f'}\left[u_x \phi_{12,u}+u_{xx}\phi_{12,u_x}-\lambda u^{2}u_x f'\pm\frac{\eta_2}{\sqrt{1+\mu_{2}^2}}\phi_{12}
-\left(2\lambda u u_x\pm\frac {\lambda \eta_2 u^{2}}{\sqrt{1+\mu_{2}^2}}\pm\frac{C}{\sqrt{1+\mu_{2}^2}}\right)f\right],
\\&\omega_{1}=f dx+(-\lambda u^{2}f+\phi_{12})dt,
\\&\omega_{2}=(\mu_{2}f+\eta_{2})dx+(-\lambda \mu_{2}u^{2}f+\mu_{2} \phi_{12}+C)dt,
\\&\omega_{3}=\left(\pm \sqrt{1+\mu_{2}^2}f\pm \frac{\mu_{2} \eta_2}{\sqrt{1+\mu_{2}^2}}\right)dx+\left(\mp \sqrt{1+\mu_{2}^2}\lambda u^{2}f\pm \sqrt{1+\mu_{2}^2}\phi_{12}\pm \frac{\mu_{2}C}{\sqrt{1+\mu_{2}^2}}\right)dt,
\end{split}
\end{equation*}
where  $\lambda ,\mu_{p}, \eta_{p}, C \in \mathbb{R}$, $(\lambda \eta_2)^{2}+C^{2} \neq 0$, $f=f(u-u_{xx})$ and $\phi_{i2}=\phi_{i2}(u,u_x)$ are real and differentiable functions satisfying $f'\neq 0$.
\end{Lemma}

Observe that Equation (1.3) belongs to the class (2.2) and by choosing
\begin{equation}
  \lambda=1, f=u-u_{xx}, \phi_{12}=-2u^2u_x+uu_x^2+u^3, C=0,\eta_2=\pm\sqrt{1+\mu_{2}^2},
\end{equation}
one can get associated one-forms $\omega_i$'s coefficient functions as
\begin{equation}
  \begin{aligned}
		f_{11} &= u - u_{xx}, & f_{12} &= u^2u_{xx} - 2u^2u_{x} + uu_x^2, \\
		f_{21} &= \mu(u - u_{xx}) \pm \sqrt{1 + \mu^2}, & f_{22} &= \mu(u^2u_{xx} - 2u^2u_{x} + uu_x^2), \\
		f_{31} &= \pm\sqrt{1 + \mu^2}(u - u_{xx}) + \mu, & f_{32} &= \pm\sqrt{1 + \mu^2}(u^2u_{xx} - 2u^2u_{x} + uu_x^2),
	\end{aligned}
\end{equation}
where $\mu$ is an arbitray real parameter.

It is not hard to check that the structure equations (2.1) are satisfied whenever $u(x,t)$ is a solution of Equation (1.3).

\begin{Theorem}
The generalized Camassa-Holm equation (1.3) describes a pseudospherical surface.  Moreover, it is geometrically integrable.
\end{Theorem}

\begin{Remark}
Let $S \subseteq \mathbb{R}^2$ be a nonempty and connected set, and suppose that $v$ is a real valued (smooth) function defined on $S$ such that $S \supseteq (x,t) \mapsto v^2v_{xx} - 2v^2v_{x} + vv_x^2$ vanishes at most on a set of measure zero. If $u(x,t)=v(x,t)$ is a solution of Equation (1.3), then
\begin{equation*}
u^2u_{xx} - 2u^2u_{x} + uu_x^2\mid_{u(x,t)=v(x,t)}=v^2v_{xx} - 2v^2v_{x} + vv_x^2\neq 0 \qquad  a.e.,
\end{equation*}
meaning that the function $v(x,t)$ is a generic solution of (1.3). In particular, a straightforward calculation shows that the generic solution for (1.3) is $u(x,t)=f(t)e^x$, where $f(\cdot)$  is a smooth function of $t$.
\end{Remark}

\section{ Pseudo-potentials and conservations laws}

\indent \indent One of the interesting properties of equations describing pseudospherical surfaces is that they admit quadratic pseudo-potentials \cite{reyes2002geometric}. Following Reyes in \cite{reyes2006pseudo}, $\Gamma$ is called a pseudo-potential for a scalar partial differential equation $\Xi=0$ if and only if $f_t=g_x$ whenever $\Gamma_x=f,\Gamma_t=g$ and $u(x,t)$ is a solution to $\Xi=0$. 

\begin{Lemma}
For a given differential equation $\Xi=0$ describing pseudospherical surfaces with associated one-forms $\omega_i$, the following Pfaffian system are completely integrable whenever $u(x,t)$ is a solution of the equation $\Xi=0$:
\begin{equation}
  -2d\Gamma=\omega_3+\omega_2-2\Gamma \omega_1+\Gamma^2(\omega_3-\omega_2)
\end{equation}
and
\begin{equation}
2d\gamma=\omega_3-\omega_2-2\gamma \omega_1+\gamma^2(\omega_3+\omega_2).
\end{equation}
Moreover, the one-forms
\begin{equation}
  \Theta=\omega_1-\Gamma(\omega_3-\omega_2)
\end{equation}
and
\begin{equation}
\hat{\Theta}=-\omega_1+\gamma(\omega_3+\omega_2)
\end{equation}
are closed whenever $u(x,t)$ is a soltion of the equation $\Xi=0$ and $\Gamma  (resp.\gamma)$ is a solution of the Pfaffian system (3.1) (resp. (3.2)).
\end{Lemma}

Geometrically, the Pfaffian systems (3.1) and (3.2) determine geodesic coordinates on the pseudospherical surfaces associated with the equation $\Xi=0$ \cite{reyes2002geometric}. The functions $\Gamma$ and $\gamma$ appearing in (3.1) and (3.2) are quadratic pseudo-potentials for the equation $\Xi=0$ describing pseudospherical surface. An easy computation using these equations shows that, conversely, if a differential equation admits a quadratic pseudo-potential then it is  of pseudospherical type \cite{reyes2011equations}. In order to apply above Lemma to Equation (1.3), it is convenient to write out $\hat{\Theta}$ as follows:
\begin{equation}
\hat{\Theta}=[\gamma (f_{21}+f_{31})-f_{11}] dx+[\gamma(f_{22}+f_{32})-f_{12}]dt.
\end{equation}

Our main result regarding pseudo-potentials for Equation (1.3) is:

\begin{Theorem}
The generalized Camassa-Holm equation (1.3) admits a quadratic pseudo-potnetial $\gamma$ determined by the equations
\begin{equation}
2\gamma_x =2\gamma+\gamma^2\eta(u-u_{xx}+1),\quad 2\gamma_t=\gamma^2\eta(u^2u_{xx}-2u^2u_x+uu_x^2),
\end{equation}
in which $\eta \neq0$ is a parameter. Moreover, Equation (1.3) possesses the parameter-dependent conservation law
\begin{equation}
[\gamma(u-u_{xx}+1)]_t=[\gamma(u^2u_{xx}-2u^2u_x+uu_x^2)]_x.
\end{equation}
\end{Theorem}

\begin{Proof}
A straightforward calculation after substituting (2.4) into (3.2) gives
\begin{equation}
\begin{aligned}
2\gamma_x &=(\pm\sqrt{1+\mu^2}-\mu)(u-u_{xx}-1)-2\gamma(u-u_{xx})\\
&+\gamma^2(\pm\sqrt{1+\mu^2}+\mu)(u^2u_{xx}-2u^2u_x+uu_x^2)
\end{aligned}
\end{equation}
and
\begin{equation}
\begin{aligned}
2\gamma_t &=(\pm\sqrt{1+\mu^2}-\mu)(u^2u_{xx}-2u^2u_x+uu_x^2)-2\gamma(u^2u_{xx}-2u^2u_x+uu_x^2)\\
&+\gamma^2(\pm\sqrt{1+\mu^2}+\mu)(u^2u_{xx}-2u^2u_x+uu_x^2).
\end{aligned}
\end{equation}
Set $\eta \equiv \mu\pm\sqrt{1+\mu^2}\neq 0$. Then we have $\eta^{-1} =- \mu\pm\sqrt{1+\mu^2}$. Replacing $\gamma$ by $\gamma+\eta^{-1}$ in (3.8) and (3.9), we get (3.6). Substituting (2.4) into (3.5), one can obtain
\begin{equation}
  [\gamma(u-u_{xx}+1)]_t=[\gamma(u^2 u_{xx}-2u^2 u_x +u u_x^2)]_x+(\pm\sqrt{1+\mu^2}-\mu)(u^2 u_{xx}-2u^2 u_x +u u_x^2).
\end{equation}
After taking the transformation for $\eta$ and $\gamma$, (3.10) can be rewritten as (3.7), in which we would use the following equivalent form of Equation (1.3)
\begin{equation}
  (u-u_{xx})_t=(u^2 u_{xx}-2u^2 u_x +u u_x^2)_x+(u^2 u_{xx}-2u^2 u_x +u u_x^2).
\end{equation}
\end{Proof}

\hspace*{\fill}\\
\indent Conservation laws for Equation (1.3) are obtained by expanding Equations (3.6) in powers of $\eta$. Two consistent expansions of $\gamma$ will be used. The first one is
\begin{equation}
  \gamma=\sum^{\infty}_{k=1} \gamma_k \eta^{-k}.
\end{equation}
Substituting (3.12) into (3.6), we arrive at the following system of $\gamma_k$:
\begin{equation}
\begin{aligned}
 \gamma_{1,x}& =\frac{1}{2} \gamma_1^2(u-u_{xx}+1)+\gamma_1, \\
 \gamma_{2,x}&=\frac{1}{2}(2 \gamma_1 \gamma_2)(u-u_{xx}+1)+\gamma_2,\\
 \gamma_{3,x}&=\frac{1}{2}(2\gamma_1\gamma_3+\gamma_2^2)(u-u_{xx}+1)+\gamma_3,\\
&\dots,\\
 \gamma_{k,x}&=\frac{1}{2}\sum_{i=1}^k \gamma_i \gamma_{k+1-i}(u-u_{xx}+1)+\gamma_k.
\end{aligned}
\end{equation}
Solving the system, we obtain the few solutions given by
\begin{equation}
\begin{aligned}
     \gamma_1 &=\frac{2}{u_x-u-1}, \\
     \gamma_2 &=\frac{1}{e^x (u_x-u-1)^2},\\
     \gamma_3 &=\frac{1}{2e^{2x} (u_x-u-1)^3},\\
     &\dots,\\
      \gamma_k &=\frac{1}{2^{k-2}e^{(k-1)x} (u_x-u-1)^k},\quad k\geq2.
\end{aligned}
\end{equation}

Next, we consider the second expansion of $\gamma$ in the form
\begin{equation}
  \gamma=\sum^{\infty}_{k=0} \Gamma_k \eta^{k}
\end{equation}
Similarly, substituting it into (3.6) yields the following system:
\begin{equation}
\begin{aligned}
 \Gamma_{0,x}&= \Gamma, \\
 \Gamma_{1,x}& =\frac{1}{2} \Gamma_0^2(u-u_{xx}+1)+\Gamma_1, \\
 \Gamma_{2,x}&=\frac{1}{2}(2 \Gamma_0 \Gamma_1)(u-u_{xx}+1)+\Gamma_2,\\
 \Gamma_{3,x}&=\frac{1}{2}(2\Gamma_0\Gamma_2+\Gamma_1^2)(u-u_{xx}+1)+\Gamma_3,\\
&\dots,\\
 \Gamma_{k,x}&=\frac{1}{2}\sum_{i=0}^{k-1} \Gamma_i \Gamma_{k-1-i}(u-u_{xx}+1)+\Gamma_k.
\end{aligned}
\end{equation}

The solutions of the equations in (3.16) are
\begin{equation}
\begin{aligned}
     \Gamma_0 &=e^x, \\
     \Gamma_1 &=\frac{1}{2}e^{2x} (u-u_x+1),\\
     \Gamma_2 &=\frac{1}{4}e^{3x} (u-u_x+1)^2,\\
     \Gamma_3 &=\frac{1}{8}e^{4x} (u-u_x+1)^3,\\
     &\dots,\\
      \Gamma_k &=\frac{1}{2^k}e^{(k+1)x} (u-u_x+1)^k.
\end{aligned}
\end{equation}

\begin{Theorem}
Let $\gamma$ given by (3.7). Then, for each positive integer $k>1$, the generalized Camassa-Holm equation (1.3) has the following conservation laws:
\begin{equation}
\left[ \frac{u-u_{xx}+1}{e^{(k-1)x}(u_x-u-1)^k}\right]_t=\left[ \frac{u^2u_{xx}-2u^2u_x+uu_x^2}{e^{(k-1)x}(u_x-u-1)^k}\right]_x.
\end{equation}
and
\begin{equation}
\left[e^{(k+1)x}(u-u_x+1)^k) (u-u_{xx}+1)\right]_t=\left[e^{(k+1)x}(u-u_x+1)^k (u^2u_{xx}-2u^2u_x+uu_x^2)\right]_x.
\end{equation}
\end{Theorem}

\begin{Lemma}
On the solutions of Equation (1.3) we have the following identity:
\begin{equation}
  (u-u_x)_t=u^2u_{xx}-2u^2u_x+uu_x^2.
\end{equation}
\end{Lemma}

\begin{Proof}
Substituting (3.12) into the second equation in (3.6) and comparing the coefficients of $\eta^{-k}$, we get
\begin{equation}
  \gamma_{k,t}=\frac{1}{2}\sum_{i=1}^k \gamma_i \gamma_{k+1-i}(u^2u_{xx}-2u^2u_x+uu_x^2).
\end{equation}
It follows from the last equation in (3.14) that
\begin{equation}
\sum_{i=1}^k \gamma_i \gamma_{k+1-i}=\frac{8k}{2^k (u_x-u-1)^{k+1}e^{(k-1)x}}
\end{equation}
and
\begin{equation}
\gamma_{k,t}=-\frac {4k(u_x-u)_t}{2^k (u_x-u-1)^{k+1} e^{(k-1)x}},\quad k\geq 1.
\end{equation}
Thus, after a simple calculation, we have (3.20).

On the other hand, when $\gamma$ is given by (3.15),  similarly we can obtain
\begin{equation}
  \sum_{i=0}^{k-1} \Gamma_i \Gamma_{k-1-i}=\frac{k}{2^{k-1}}e^{(k+1)x}(u-u_x+1)^{k-1},
\end{equation}
and
\begin{equation}
\Gamma_{k,t}=\frac{k}{2^k}e^{(k+1)x}(u-u_x+1)^{k-1}(u-u_x)_t, \quad k\geq1,
\end{equation}
Then one also can have (3.20).
\end{Proof}

\begin{Theorem}
Let $u(x,t)$ be a solution of  the generalized Camassa-Holm equation (1.3). Then the two one-forms
\begin{equation}
\varpi_1=\frac {u-u_{xx}+1}{e^{(k-1)x}(u_x-u-1)^k}dx+\frac{u^2u_{xx}-2u^2u_x+uu_x^2}{e^{(k-1)x}(u_x-u-1)^k} dt,\quad k\geq 2
\end{equation}
and
\begin{equation}
\begin{aligned}
\varpi_2&=e^{(k+1)x}(u-u_x+1)^k(u-u_{xx}+1)dx \\
&+e^{(k+1)x}(u-u_x+1)^k (u^2u_{xx}-2u^2u_x+uu_x^2)dt,\quad k\geq1
\end{aligned}
\end{equation}
are exact. In particular, the conservation laws (3.18) and (3.19) are trivial.
\end{Theorem}

\begin{Proof}
It only needs to prove that there exists two function $\phi_{jk}\in \mathcal{A}$ such that $\varpi_{jk}=d\phi_{jk}=\partial_x\phi_{jk}dx+\partial_t \phi_{jk}dt$, where $j=1,2$. A straightforward calculation show that
\begin{equation*}
  \frac{u-u_{xx}+1}{e^{(k-1)x}(u_x-u-1)^k}=\partial_x \frac{1}{(k-1)e^{(k-1)x}(u_x-u-1)^{k-1}},
\end{equation*}
and
\begin{equation*}
  e^{(k+1)x}(u-u_x+1)^k(u-u_{xx}+1)=\partial_x \frac{e^{(k+1)x}(u-u_x+1)^{k+1}}{k+1}.
\end{equation*}
Define
\begin{equation*}
  f_{1k}=\frac{1}{(k-1)e^{(k-1)x}(u_x-u-1)^{k-1}},\quad  f_{2k}=\frac{e^{(k+1)x}(u-u_x+1)^{k+1}}{k+1},
\end{equation*}
it is not difficult to get
\begin{equation*}
\partial_t f_{1k}=\frac{(u-u_x)_t}{e^{(k-1)x}(u_x-u-1)^k}=\frac{u^2u_{xx}-2u^2u_x+uu_x^2}{e^{(k-1)x}(u_x-u-1)^k},
\end{equation*}
and
\begin{equation*}
\partial_t f_{2k}=e^{(k+1)x}(u-u_x+1)^k (u-u_x)_t=e^{(k+1)x}(u-u_x+1)^k(u^2u_{xx}-2u^2u_x+uu_x^2),
\end{equation*}
in which we have used Lemma 3. The results are proved by taking $\phi_{jk}=f_{jk},j=1,2$.
\end{Proof}

\section{ Local isometric immersions}

\indent \indent In a series of recent works \cite{castrosilva2016third, kahouadji2015local, kahouadji2013second, kahouadji2019local, freire2022novikov, ferraioli2022isometric}, the problem of local isometric immersions has been investigated for the families of equations describing pseudospherical surfaces. Here we consider local isometric immersions of the pseudospherical surfaces described by generic solutions of the generalized Camassa-Holm equation (1.3).  We recall some basic facts and notations from the classical theory of surfaces. The interested reader is referred to \cite{ivey2003cartan} for further details.

Introducing the one-forms
\begin{equation}\label{4.1}
\omega_{13}=a\omega_1+b\omega_2, \quad \omega_{23}=b\omega_1+c\omega_2,
\end{equation}
with $a,b,c$ differentiable functions depending on $x,t,u$ and the derivatives of $u$, $\omega_1, \omega_2,\omega_3$ one-forms given by (1.4) and (2.4). Then the Gauss-Codazzi equations read
\begin{equation}\label{4.2}
ac-b^2=-1,
\end{equation}
\begin{equation}\label{4.3}
d\omega_{13}=\omega_{12}\wedge\omega_{23}, \quad d\omega_{23}=\omega_{21}\wedge\omega_{13},
\end{equation}
and the second fundamental form of the corresponding pseudospherical surfaces writes as
\begin{equation}\label{4.4}
\Pi=\omega_1 \cdot \omega_{13}+\omega_2 \cdot \omega_{23},
\end{equation}
while the mean curvature of the isometric immersion is given by
\begin{equation}\label{4.5}
H=\frac{a+c}{2}.
\end{equation}

It follows from \cite{kahouadji2013second} that the Codazzi equations (4.3) may be expressed in terms of the components $f_{ij}$ of the one-forms $\omega_1,\omega_2,\omega_3$ in the form
 \begin{equation}\label{4.6}
 f_{11}D_t a+f_{21}D_t b-f_{12}D_x a-f_{22}D_x b-2b\Delta_{13}+(a-c)\Delta_{23}=0,
 \end{equation}
\begin{equation}\label{4.7}
f_{11}D_t b+f_{21}D_t c-f_{12}D_x b-f_{22}D_x c+(a-c)\Delta_{13}+2b\Delta_{23}=0,
\end{equation}
where
\begin{equation}\label{4.8}
\Delta_{ij}=f_{i1}f_{j2}-f_{j1}f_{i2}
\end{equation}
and where we assume that
\begin{equation}\label{4.9}
\Delta_{13}^2+\Delta_{23}^2\neq 0.
\end{equation}

In view of Bonnet theorem, the local isometric immersion of the pseudospherical surfaces depicted by any generic solution $u(x,t)$ of a equation describing pseudospherical surfaces exists if and only if there ia always a triple $\{a,b,c\}$ of differentiable functions satisfying the Gauss-Codazzi equations and its second fundamental form is written as (4.4). However, in general the dependence of $\{a,b,c\}$ on $u(x,t)$ may be quite complicated and the works reported \cite{ kahouadji2015local, kahouadji2013second, kahouadji2019local, ferraioli2022isometric} so far suggest that, except for the sine-Gordon equation and the short-pulse equation and some generalizations, almost all the equations describing pseudospherical surfaces admitted only local isometric immersions with ''universal" triples $\{a,b,c\}$, depengding only on $x$ and $t$.

\begin{Theorem}\label{7}
Let $u(x,t)$ be any generic solution of the generalized Camassa-Holm equation (1.3), with $a,b,c$ the coefficients of the second fundamental form depending on $x,t,u$ and the derivatives of $u$. There exists a local isometric immersion in $\mathbb{R}^2$ of a pseudospherical surfaces if and only if

$(i) \quad \mu=0$ and $a,b$ and $c$ depend only on $x$ and are given by
\begin{equation}\label{4.10}
a=\pm \sqrt{L(x)}, \quad b=-\beta e^{2x}, \quad c=a-a',
\end{equation}
where $L(x)=C e^{2x}-\beta^2 e^{4x}-1$, with $C,\beta \in \mathbb{R},C>0,C^2>4\beta^2$ and the one-forms $\omega_1,\omega_2,\omega_3$, given by (1.4), are defined on a strip of $\mathbb{R}^2$ where
\begin{equation}\label{4.11}
log \sqrt{\frac{C-\sqrt{C^2-4\beta^2}}{2\beta^2}}<x<log \sqrt{\frac{C+\sqrt{C^2-4\beta^2}}{2\beta^2}}.
\end{equation}
Moreover, the constant $C$ and $\beta$ are chosen so that the strip intersects the domain of the solution of Equation (1.3).

$(ii) \quad \mu \neq0$ and $a,b$ and $c$ depend only on $x$ and are given by
\begin{equation}\label{4.12}
  \begin{aligned}
    a&= \frac{1}{2\mu}[\pm\mu\sqrt{\Delta}-(\mu^2 -1)b+\beta e^{2x}],\\
    c&= \frac{1}{2\mu}[\pm\mu\sqrt{\Delta}+(\mu^2 -1)b-\beta e^{2x}],\\
    \Delta &=\frac{[(\mu^2-1)b-\beta e^{2x}]^2-4\mu^2(1-b^2)}{\mu^2}>0,
  \end{aligned}
\end{equation}
where $b$ satisfies the ordinary differential equation
\begin{equation}\label{4.13}
\begin{split}
 & [\pm(\mu^2+1)^2 b\mp(\mu^2 -1)\beta e^{2x}+\mu(\mu^2 +1)\sqrt{\Delta}]b' \\
 & +2[(-\mu(\mu^2+1)\sqrt{\Delta}\mp(\mu^2-1)\beta e^{2x})b\pm\beta^2 e^{4x}]=0.
\end{split}
\end{equation}
\end{Theorem}

\begin{Proof}
First, we will show that $a,b$ and $c$ depend only on $x$ and $t$. Then we are going to determine such coefficients for the generalized Camassa-Holm equation (1.3).

If the coefficients of the second fundamental form depending on $x,t,u$ and the derivatives of $u$ and the dunctions $f_{ij}$ only depend on $(u_0,u_1,u_2,\cdots, u_k)$, then (4.6) and (4.7) rewrites as
\begin{equation}\label{4.14}
\begin{split}
&f_{11}a_t+f_{21}b_t-f_{12}a_x-f_{22}b_x-2b(f_{11}f_{32}-f_{12}f_{31})+(a-c)(f_{21}f_{32}-f_{22}f_{31}) \\
&-\sum_{i=0}^l(f_{12}a_{u_i}+f_{22}b_{u_i})u_{i+1}+\sum_{i=0}^l(f_{11}a_{u_i}+f_{21}b_{u_i})u_{i,t} \\
&-\sum_{j=0}^m(f_{12}a_{w_j}+f_{22}b_{w_j})w_{j,x}+\sum_{j=0}^m(f_{11}a_{w_j}+f_{21}b_{w_j})w_{j+1} \\
&-\sum_{k=0}^n(f_{12}a_{v_k}+f_{2}b_{v-k})v_{k,x}+\sum_{k=0}^n(f_{11}a_{v_k}+f_{21}b_{v_k})v_{k+1}=0,
\end{split}
\end{equation}
and
\begin{equation}\label{4.15}
\begin{split}
&f_{11}b_t+f_{21}c_t-f_{12}b_x-f_{22}c_x+(a-c)(f_{11}f_{32}-f_{12}f_{31})+2b(f_{21}f_{32}-f_{22}f_{31}) \\
&-\sum_{i=0}^l(f_{12}b_{u_i}+f_{22}c_{u_i})u_{i+1}+\sum_{i=0}^l(f_{11}b_{u_i}+f_{21}c_{u_i})u_{i,t} \\
&-\sum_{j=0}^m(f_{12}b_{w_j}+f_{22}c_{w_j})w_{j,x}+\sum_{j=0}^m(f_{11}b_{w_j}+f_{21}c_{w_j})w_{j+1} \\
&-\sum_{k=0}^n(f_{12}b_{v_k}+f_{2}c_{v-k})v_{k,x}+\sum_{k=0}^n(f_{11}b_{v_k}+f_{21}c_{v_k})v_{k+1}=0,
\end{split}
\end{equation}
where $u_i=\partial_x^i u, w_j=\partial_t^j u,v_k=\partial_x^k u_x$, $1\leq l,m,n\leq \infty $ are finite, but otherwise arbitrary.

Without loss of generality, we assume $m=n$. In fact, suppose the case $m<n$, since $n\geq m+1$, differentiating (4.14), (4.15) and (4.2) with respect to $v_{n+1}$ leads to $a_{v_n}=b_{v_n}=c_{v_n}=0$. Successive differentiation with respect to $v_n,v_{n-1},\cdots,v_{(m+1)+1}$ leads to $a_{v_{n-1}}=\cdots=a_{v_{m+1}}=0, b_{v_{n-1}}=\cdots=b_{v_{m+1}}=0, c_{v_{n-1}}=\cdots =c_{v_{m+1}}=0$. Thus, $a,b$ and $c$ are functions of $x,t,u,u_1,\cdots, u_i, w_1,\cdots,w_m,v_1,\cdots, v_m$. For the case $m<n$, since $m\geq n+1$, similarly, successive differentiation with respect to $w_{m+1},v_m,\cdots,v_{(n+1)+1}$ leads to $a_{w_m}=\cdots=a_{w_{n+1}}=0, b_{w_m}=\cdots=b_{w_{n+1}}=0, c_{w_m}=\cdots =c_{w_{n+1}}=0$. Thus, $a,b$ and $c$ are functions of $x,t,u,u_1,\cdots, u_i, w_1,\cdots,$ $w_n,v_1,\cdots, v_n$.

In views of (2.4), by deriving equations (4.14) and (4.15) with respect to $v_{n+1}$ and $w_{n+1}$, one gets
\begin{equation}\label{4.16}
\begin{split}
&b_{v_n}= -\frac{f_{11}}{f_{21}} a_{v_n},\quad b_{w_n}=-\frac{f_{11}}{f_{21}} a_{w_n},\\
&c_{v_n}= \left(\frac{f_{11}}{f_{21}}\right)^2 a_{v_n},\quad c_{w_n}=\left(\frac{f_{11}}{f_{21}}\right)^2 a_{w_n},
\end{split}
\end{equation}
then the derivative of Gauss equation (4.2) with respect to $v_n$ and $w_n$ returns
\begin{equation}\label{4.17}
\left[c+\left(\frac{f_{11}}{f_{21}}\right)^2 a+2b\frac{f_{11}}{f_{21}}\right]a_{v_n}=0,\quad \left[c+\left(\frac{f_{11}}{f_{21}}\right)^2 a+2b\frac{f_{11}}{f_{21}}\right]a_{w_n}=0,
\end{equation}
and we will proceed by further distinguishing the two cases:

\hspace*{\fill}\\
$(i) \quad c+\left(\frac{f_{11}}{f_{21}}\right)^2 a+2b\frac{f_{11}}{f_{21}}\neq 0;$

\hspace*{\fill}\\
$(ii) \quad c+\left(\frac{f_{11}}{f_{21}}\right)^2 a+2b\frac{f_{11}}{f_{21}}=0$.

\hspace*{\fill}\\
\noindent\textbf{Case (i).} Suppose $l=1$. It follows from (4.16) and (4.17) that $a_{v_n}=a_{w_n}=0,b_{v_n}=b_{w_n}=0$ and $c_{v_n}=c_{w_n}=0$. By successive differentiation of (4.14), (4.15) and (4.2) with respect to $v_n,v_{n-1},\cdots v_1$ and $w_n,w_{n-1},\cdots w_1$, one would get that $a_{v_k}=b_{v_k}=c_{v_k}=0$ and $a_{w_k}=b_{w_k}=c_{w_k}=0$ for $k=0,1,\cdots,n$. Therefore, $a,b$ and $c$ depend only on $x$ and $t$.

Suppose $l \geq 2$. Successive differentiation of (4.14), (4.15) and (4.2) with respect to $v_{n+1},v_n,\cdots v_2$ and $w_{n+1},w_n,\cdots w_2$ leads to $a_{v_k}=b_{v_k}=c_{v_k}=0$ and $a_{w_k}=b_{w_k}=c_{w_k}=0$ for $k=1,2,\cdots,n$. Thus, $a,b$ and $c$ depend on $x,t,u_0=w_0,u_1=v_0,\cdots, u_l$. And the equations (4.14) and (4.15) are equivalent to
\begin{equation}\label{4.18}
\begin{split}
&f_{11}a_t+f_{21}b_t-f_{12}a_x-f_{22}b_x-2b(f_{11}f_{32}-f_{12}f_{31})+(a-c)(f_{21}f_{32}-f_{22}f_{31}) \\
&-\sum_{i=0}^l(f_{12}a_{u_i}+f_{22}b_{u_i})u_{i+1}+\sum_{i=2}^l(f_{11}a_{u_i}+f_{21}b_{u_i})\partial_x^{i-2}(u_{0,t}-F) \\
&+(f_{11}a_{w_0}+f_{21}b_{w_0})w_{1}+(f_{11}a_{v_0}+f_{21}b_{v_0})v_{1}=0,
\end{split}
\end{equation}
and
\begin{equation}\label{4.19}
\begin{split}
&f_{11}b_t+f_{21}c_t-f_{12}b_x-f_{22}c_x+(a-c)(f_{11}f_{32}-f_{12}f_{31})+2b(f_{21}f_{32}-f_{22}f_{31}) \\
&-\sum_{i=0}^l(f_{12}b_{u_i}+f_{22}c_{u_i})u_{i+1}+\sum_{i=2}^l(f_{11}b_{u_i}+f_{21}c_{u_i})\partial_x^{i-2}(u_{0,t}-F) \\
&+(f_{11}b_{w_0}+f_{21}c_{w_0})w_{1}+(f_{11}b_{v_0}+f_{21}c_{v_0})v_{1}=0,
\end{split}
\end{equation}
where $F=u_0^2 u_3 - u_0^2u_2-3u_0u_1^2-2u_0^2u_1+4u_0 u_1 u_2+u_1^3$ in views of the generalized Camassa-Holm equation (1.3).

Differentiating (4.18) and (4.19) with respect to $u_{l+1}$ respectively, we have
\begin{equation}\label{4.20}
\phi_{12}a_{u_l}+\phi_{22}b_{u_l}=0,\quad \phi_{12}b_{u_l}+\phi_{22}c_{u_l}=0
\end{equation}
with $\phi_{12}=-2u_0^2u_1+u_0u_1^2+u_0^3,\phi_{22}=\mu\phi_{12}\pm\sqrt{1+\mu^2}u_0^2,\mu\in\mathbb{R}$ and $f_{i2}=-u_0^2f_{i1}+\phi_{i2}$.
Since $\phi_{22}\neq0$ on a non-empty open set, from (4.20), we obtain
\begin{equation}\label{4.21}
b_{u_l}=-\frac{\phi_{12}}{\phi_{22}}a_{u_l},\quad c_{u_l}=\left(\frac{\phi_{12}}{\phi_{22}}\right)^2a_{u_l}.
\end{equation}
Differentiating the Gauss equation (4.2) with respect to $u_l$ leads to $a_{u_l}c+ac_{u_l}-2bb_{u_l}=0$, which implies using (4.20) that
\begin{equation}\label{4.22}
\left[c+\left(\frac{\phi_{12}}{\phi_{22}}\right)^2 a+2b\frac{\phi_{12}}{\phi_{22}}\right]a_{u_l}=0,
\end{equation}

If $c+\left(\frac{\phi_{12}}{\phi_{22}}\right)^2 a+2b\frac{\phi_{12}}{\phi_{22}}=0$,  then it follows from the Gauss equation (4.2) that
\begin{equation}\label{4.23}
b=\pm1-\frac{\phi_{12}}{\phi_{22}}a, \quad c=\left(\frac{\phi_{12}}{\phi_{22}}\right)^2 a\mp2\frac{\phi_{12}}{\phi_{22}}
\end{equation}
and the following identities hold:
\begin{equation}\label{4.24}
\begin{split}
&D_t b=-\frac{\phi_{12}}{\phi_{22}}D_t a-aD_t \left(\frac{\phi_{12}}{\phi_{22}}\right), D_t c=\left(\frac{\phi_{12}}{\phi_{22}}\right)^2 D_t a+2 \left(\frac{\phi_{12}}{\phi_{22}}a\mp1\right)D_t \left(\frac{\phi_{12}}{\phi_{22}}\right),\\
&D_x b=-\frac{\phi_{12}}{\phi_{22}}D_x a-aD_x \left(\frac{\phi_{12}}{\phi_{22}}\right), D_x c=\left(\frac{\phi_{12}}{\phi_{22}}\right)^2 D_x a+2 \left(\frac{\phi_{12}}{\phi_{22}}a\mp1\right)D_x \left(\frac{\phi_{12}}{\phi_{22}}\right),
\end{split}
\end{equation}
where $D_t$ and $D_x$ are total derivative operators. Therefore, by using (4.24), equations (4.6) becomes
\begin{equation}\label{4.25}
\frac{\Delta_{12}}{\phi_{22}}(D_t a+u_0^2D_x a)-af_{21}D_t \left(\frac{\phi_{12}}{\phi_{22}}\right)+af_{22}D_x \left(\frac{\phi_{12}}{\phi_{22}}\right)-2b\Delta_{13}+(a-c)\Delta_{23}=0
\end{equation}
and (4.7) becomes
\begin{equation}\label{4.26}
\begin{split}
&-\frac{\phi_{12}}{\phi_{22}}\frac{\Delta_{12}}{\phi_{22}}(D_t a+u_0^2D_x a)+\left[-f_{11}a+2f_{21}\left(\frac{\phi_{12}}{\phi_{22}}\mp1\right)\right]D_t \left(\frac{\phi_{12}}{\phi_{22}}\right)\\
&-\left[a\left(u_0^2 \frac{\Delta_{12}}{\phi_{22}}+f_{22}\frac{\phi_{12}}{\phi_{22}}\right)\mp2f_{22}\right]D_x \left(\frac{\phi_{12}}{\phi_{22}}\right)+(a-c)\Delta_{13}+2b\Delta_{23}=0.
\end{split}
\end{equation}
Adding (4.25) multiplied by $\frac{\phi_{12}}{\phi_{22}}$ to (4.26) we get
\begin{equation}\label{4.27}
\begin{split}
&\left(-\frac{\Delta_{12}}{\phi_{22}}a \mp 2 f_{21}\right)\left[\left(\frac{\phi_{12}}{\phi_{22}}\right)_{u_0} w_1+\left(\frac{\phi_{12}}{\phi_{22}}\right)_{u_1}v_1\right]-\left[a u_0^2 \frac{\Delta_{12}}{\phi_{22}}\mp2f_{22}\right]D_x \left(\frac{\phi_{12}}{\phi_{22}}\right)\\
&+\left(a-c-2b\frac{\phi_{12}}{\phi_{22}}\right)\Delta_{13}+\left[\frac{\phi_{12}}{\phi_{22}}(a-c)+2b\right]\Delta_{23}=0.
\end{split}
\end{equation}
Differentiating (4.27) with respect to $v_1$ and $w_1$ respectively, we get
\begin{equation}\label{4.28}
\left(-\frac{\Delta_{12}}{\phi_{22}}a \mp 2 f_{21}\right)\left(\frac{\phi_{12}}{\phi_{22}}\right)_{u_0}=0,\quad  \left(-\frac{\Delta_{12}}{\phi_{22}}a \mp 2 f_{21}\right)\left(\frac{\phi_{12}}{\phi_{22}}\right)_{u_1}=0,
\end{equation}
It is obvious that $-\frac{\Delta_{12}}{\phi_{22}}a \mp 2 f_{21}=0$ on an non-empty open set, then $a=\mp2 \frac{\phi_{22}}{f_{21}}\Delta_{12}$. However, using such $a$ and (4.23) we obtain
\begin{equation*}
c+\left(\frac{f_{11}}{f_{21}}\right)^2 a+2b\frac{f_{11}}{f_{21}}=0,
\end{equation*}
which contradicts the hypothesis.

Therefore, $c+\left(\frac{\phi_{12}}{\phi_{22}}\right)^2 a+2b\frac{\phi_{12}}{\phi_{22}}\neq0$ and by (4.22) we have $a_{u_l}=0$  and thus, by (4.21), $b_{u_l}=c_{u_l}=0$. Successive differentiation of (4.28), (4.19) and (4.2) with respect to $u_{l-1},\cdots,u_3$ leads to $a_{u_{l-1}}=a_{u_{l-2}}=\cdots=a_{u_2}=0$ and then $b_{u_{l-1}}=b_{u_{l-2}}=\cdots=b_{u_2}=0$ and $c_{u_{l-1}}=c_{u_{l-2}}=\cdots=c_{u_2}=0$. Hence, equations (4.18) and (4.19) rewrites as respectively
\begin{equation}\label{4.29}
\begin{split}
&f_{11}a_t+f_{21}b_t-f_{12}a_x-f_{22}b_x-2b(f_{11}f_{32}-f_{12}f_{31})+(a-c)(f_{21}f_{32}-f_{22}f_{31}) \\
&-\sum_{i=0}^1(f_{12}a_{u_i}+f_{22}b_{u_i})u_{i+1}+(f_{11}a_{w_0}+f_{21}b_{w_0})w_{1}+(f_{11}a_{v_0}+f_{21}b_{v_0})v_{1}=0,
\end{split}
\end{equation}
and
\begin{equation}\label{4.30}
\begin{split}
&f_{11}b_t+f_{21}c_t-f_{12}b_x-f_{22}c_x+(a-c)(f_{11}f_{32}-f_{12}f_{31})+2b(f_{21}f_{32}-f_{22}f_{31}) \\
&-\sum_{i=0}^1(f_{12}b_{u_i}+f_{22}c_{u_i})u_{i+1}+(f_{11}b_{w_0}+f_{21}c_{w_0})w_{1}+(f_{11}b_{v_0}+f_{21}c_{v_0})v_{1}=0.
\end{split}
\end{equation}
By deriving equations (4.29) and (4.30) with respect to $w_1$ and $v_1$ respectively, one gets
\begin{equation}\label{4.31}
\begin{split}
&b_{w_0}=-\frac{f_{11}}{f_{21}}a_{w_0},\quad c_{w_0}=\left(\frac{f_{11}}{f_{21}}\right)^2a_{w_0},\\
&b_{v_0}=-\frac{f_{11}}{f_{21}}a_{w_0},\quad c_{v_0}=\left(\frac{f_{11}}{f_{21}}\right)^2a_{w_0}.
\end{split}
\end{equation}
Thus the derivatives of Gauss equation (4.2) with respect to $w_0$ and $v_0$ return
\begin{equation*}
\left[c+\left(\frac{f_{11}}{f_{21}}\right)^2 a+2b\frac{f_{11}}{f_{21}}\right]a_{w_0}=0,\quad \left[c+\left(\frac{f_{11}}{f_{21}}\right)^2 a+2b\frac{f_{11}}{f_{21}}\right]a_{v_0}=0,
\end{equation*}
and we finally have $a_{w_0}=a_{v_0}=0$ and thus, by (4.31), $b_{w_0}=b_{v_0}=0$ and $c_{w_0}=c_{v_0}=0$. Hence, $a,b$ and $c$ are functions of $x$ and $t$.

\hspace*{\fill}\\
\noindent\textbf{Case (ii).} First, let us use the Gauss equation (4.2) in order to obtain $b$ and $c$ in terms of $a,f_{11}$ and $f_{22}$. We will then substitute the total derivatives of $b$ and $c$ back into (4.6) and (4.7).

If the expression between brackets in (4.17) vanishes, then Gauss equation provides
\begin{equation}\label{4.32}
b=\pm1-\frac{f_{11}}{f_{21}}a, \quad c=\left(\frac{f_{11}}{f_{21}}\right)^2 a\mp2\frac{f_{11}}{f_{21}}.
\end{equation}
and the following identities hold:
\begin{equation}\label{4.33}
\begin{split}
&D_t b=-\frac{f_{11}}{f_{21}}D_t a-aD_t \left(\frac{f_{11}}{f_{21}}\right), D_t c=\left(\frac{f_{11}}{f_{21}}\right)^2 D_t a+2 \left(\frac{f_{11}}{f_{21}}a\mp1\right)D_t \left(\frac{f_{11}}{f_{21}}\right),\\
&D_x b=-\frac{f_{11}}{f_{21}}D_x a-aD_x \left(\frac{f_{11}}{f_{21}}\right), D_x c=\left(\frac{f_{11}}{f_{21}}\right)^2 D_x a+2 \left(\frac{f_{11}}{f_{21}}a\mp1\right)D_x \left(\frac{f_{11}}{f_{21}}\right),
\end{split}
\end{equation}
where $D_t$ and $D_x$ are total derivative operators.
Moreover, using (2.4) we can see that $\left(\frac{f_{11}}{f_{21}}\right)_{u_0}+\left(\frac{f_{11}}{f_{21}}\right)_{u_2}=0$. Then equation (4.6) becomes
\begin{equation}\label{4.34}
a[f_{21}G+(u_0^2f_{21}+f_{22})u_3-f_{22}u_1]\left(\frac{f_{11}}{f_{21}}\right)_{u_2}+\frac{\Delta_{12}}{f_{21}}D_x a-2b\Delta_{13}+(a-c)\Delta_{23}=0,
\end{equation}
and (4.7) becomes
\begin{equation}\label{4.35}
\begin{split}
&\left\{-\frac{1}{f_{21}}(af_{11}\mp2f_{21})[f_{21}G+(u_0^2f_{21}+f_{22})u_3-f_{22}u_1]+a(u_1-u_3)\frac{\Delta_{12}}{f_{21}}\right\}\left(\frac{f_{11}}{f_{21}}\right)_{u_2}\\
&-\frac{f_{11}}{f_{21}}\frac{\Delta_{12}}{f_{21}}D_x a+(a-c)\Delta_{13}+2b\Delta_{23}=0.
\end{split}
\end{equation}

Next, we plan to transform expressions (4.34) and (4.35) to an equivalent form that will be more convenient to deal with. Adding equations (4.34) multiplied by $\frac{f_{11}}{f_{21}}$ with (4.35), one gets
\begin{equation}\label{4.36}
\begin{split}
&\left[\pm2f_{21}G\pm2(u_0^2f_{21}+f_{22})u_3\mp2f_{22}u_1+a(u_1-u_3)\frac{\Delta_{12}}{f_{21}}\right]\left(\frac{f_{11}}{f_{21}}\right)_{u_2}\\
&+\left[1+\left(\frac{f_{11}}{f_{21}}\right)^2\right]\left[a\Delta_{13}+\left(-\frac{f_{11}}{f_{21}}a\pm2\right)\Delta_{23}\right]=0.
\end{split}
\end{equation}
Taking the $v_k$ and $w_j$, $1\leq k\leq n,1\leq j\leq m$, derivatives of (4.36), we obtain, respectively,
\begin{equation}\label{4.37}
Qa_{v_k}=0,\quad Qa_{w_j}=0,
\end{equation}
where
\begin{equation}\label{4.38}
Q=(u_1-u_3)\frac{\Delta_{12}}{f_{21}}\left(\frac{f_{11}}{f_{21}}\right)_{u_2}+\left[1+\left(\frac{f_{11}}{f_{21}}\right)^2\right]\left(\Delta_{13}-\frac{f_{11}}{f_{21}}\Delta_{23}\right).
\end{equation}
It is obvious that $Q\neq0$ on a non-empty open set. Then, in views of (4.37), we conclude that $a_{v_k}=a_{w_j}=0, k=1,2,\cdots,n, j=1,2,\cdots, m$. Hence, $a$ depends only on $x,t,u_0,u_1,\cdots,u_l$. By differentiating (4.36) with respect to $u_l,l\geq4$, one would get that $a_{u_l}=0,l\geq4$. Moreover, the derivative of (4.34) with respect to $u_4$ returns $a_{u_3}=0$. Taking the $u_3$ derivative of (4.36), we have
\begin{equation}\label{4.39}
 \left[\pm2(u_0^2f_{21}+f_{22})-a \frac{\Delta_{12}}{f_{21}}\right]\left(\frac{f_{11}}{f_{21}}\right)_{u_2}=0.
\end{equation}
In views of (2.4) one concludes that the expression between brackets in (4.39) does vanish, i.e.
\begin{equation*}
\pm2(u_0^2f_{21}+f_{22})-a \frac{\Delta_{12}}{f_{21}}=0,
\end{equation*}
then we get
\begin{equation}\label{4.40}
a=\pm2\frac{f_{21}}{\Delta_{12}}\phi_{22}.
\end{equation}
By substituting (4.40) into (4.36), one has
\begin{equation}\label{4.41}
\pm2(f_{21}G+u_0^2u_1f_{21})\left(\frac{f_{11}}{f_{21}}\right)_{u_2}\pm2f_{21}\phi_{32}\left[1+\left(\frac{f_{11}}{f_{21}}\right)^2\right]=0,
\end{equation}
then, it follows from (4.41) that
\begin{equation}\label{4.42}
G=-\frac{1}{L}\frac{f_{21}^2+f_{11}^2}{f_{21}^2}\phi_{32}-u_0^2u_1,\quad L=\left(\frac{f_{11}}{f_{21}}\right)_{u_2}\quad \phi_{32}=\pm\sqrt{1+\mu^2}\phi_{12}+\mu u_0^2.
\end{equation}

Let us back to the generalized Camassa-Holm equation (1.3) and its associated one-forms' coefficient function (2.4). From (4.41) we get $L=\frac{\pm\sqrt{1+\mu^2}}{[\mu(u_0-u_2)\pm\sqrt{1+\mu^2}]^2}$ and
\begin{equation}\label{4.43}
\begin{split}
&-u_0^2u_2-3u_0u_1^2-u_0^2u_1+4u_0u_1u_2+u_1^3\\
&=\frac{[\mu(u_0-u_2)\pm\sqrt{1+\mu^2}]^2+(u_0-u_2)^2}{\sqrt{1+\mu^2}}[\sqrt{1+\mu^2}(-2u_0^2u_1+u_0u_1^2+u_0^3)\pm\mu u_0^2].
\end{split}
\end{equation}
Differentiating (4.43) with respect to $u_0$ and $u_2$ and adding both results lead to
\begin{equation}\label{4.44}
\begin{split}
&-2u_0u_2-3u_1^2+2u_0u_1+4u_1u_2-u_0^2\\
&=\frac{[\mu(u_0-u_2)\pm\sqrt{1+\mu^2}]^2+(u_0-u_2)^2}{\sqrt{1+\mu^2}}[\sqrt{1+\mu^2}(-4u_0^2u_1+u_1^2+3u_0^2)\pm2\mu u_0].
\end{split}
\end{equation}
By iterating above procedure twice, one would get that
\begin{equation}\label{4.45}
[\mu(u_0-u_2)\pm\sqrt{1+\mu^2}]^2+(u_0-u_2)^2=-1,
\end{equation}
which is a contradiction!

From now on, we are going to determine the coefficients of the second fundamental form of such local isometric immersion.  Since  $a,b$ and $c$ are universal,  then (4.6) and (4.7) become
\begin{equation}\label{4.46}
f_{11} a_t+f_{21}b_t-f_{12}a_x-f_{22}b_x-2b\Delta_{13}+(a-c)\Delta_{23}=0,
\end{equation}
\begin{equation}\label{4.47}
f_{11} b_t+f_{21}c_t-f_{12}b_x-f_{22}c_x+(a-c)\Delta_{13}+2b\Delta_{23}=0,
\end{equation}
where $\Delta_{13}=\mu f_{11}u_0^2-\mu\phi_{12}$ and $\Delta_{23}=\phi_{12}-u_0^2 f_{11}(\neq 0)$. Since $f_{ij}$ are given by (2.4), then (4.46) and (4.47) rewrites as
\begin{equation}\label{4.48}
f_{11}[a_t+\mu b_t+u_0^2 a_x+\mu u_0^2 b_x-(a-c+2\mu b)u_0^2]-\phi_{12}[a_x+\mu b_x-(a-c-2\mu b)]\pm \sqrt{1+\mu^2}b_t=0,
\end{equation}
\begin{equation}\label{4.49}
f_{11}[b_t+\mu c_t+u_0^2 b_x+\mu u_0^2 c_x+(\mu a-\mu c- 2b)u_0^2]-\phi_{12}[b_x+\mu c_x+\mu a-\mu c-2b]\pm \sqrt{1+\mu^2}c_t=0.
\end{equation}
Differentiating (4.48) and (4.49) with respect to $u_2$, one gets
\begin{equation}\label{4.50}
a_t+\mu b_t+u_0^2 a_x+\mu u_0^2 b_x-(a-c+2\mu b)u_0^2=0,
\end{equation}
\begin{equation}\label{4.51}
b_t+\mu c_t+u_0^2 b_x+\mu u_0^2 c_x+(\mu a-\mu c- 2b)u_0^2=0.
\end{equation}
Considering the coefficient of $u_0^2$ and substituting the result back into (4.50) and (4.51) lead to
\begin{equation}\label{4.52}
a_t+\mu b_t=0,\quad a_x+\mu b_x-(a-c+2\mu b)=0,
\end{equation}
\begin{equation}\label{4.53}
b_t+\mu c_t=0,\quad b_x+\mu c_x+(\mu a-\mu c- 2b)=0.
\end{equation}
In views of (4.48) and (4.49), we conclude $a_t=b_t=c_t=0$. Hence, $a,b$ and $c$ are functions depending only on $x$. From the both second equation of (4.52) and (4.53), we have
\begin{equation}\label{4.54}
\mu(-a_{xx}+2a_x)+(1+\mu^2)(b_x-2b)+\mu^2(-b_{xx}+2b_x)=0,
\end{equation}
that is
\begin{equation}\label{4.55}
\mu a_x=(1+\mu^2)b-\mu^2 b_x+\beta e^{2x},
\end{equation}
where $\beta$ is a constant.

If $\mu=0$, then it follows from (4.55) and (4.52) that
\begin{equation}\label{4.56}
b=-\beta e^{2x},\quad c=a-a_x.
\end{equation}
Substituting such $b$ and $c$ in the Gauss equation (4.2) would produce $a=\pm\sqrt{Z(x)}$ where $Z(x)=Ce^{2x}-\beta^2 e^{4x}-1$, with $C,\beta \in \mathbb{R}, C>0$ and $C^2 > 4\beta^2$. This $a$ with (4.56) gives us (4.10), where $a$ is defined on the strip described by (4.11).

If $\mu \neq0$, then (4.55) gives us $a_x$, which substituted into the second equation of (4.52) implied that
\begin{equation}\label{4.57}
c=a+\varphi,\quad \varphi=\varphi (x)=(\mu-\frac{1}{\mu})b-\frac{\beta}{\mu}e^{2x}.
\end{equation}
Substituting (4.57) into the Gauss equation we have $a^2+a\varphi-b^2=-1$, which resolved as a second order equation in terms of $a$ leads to
\begin{equation}\label{4.58}
a=\frac{-\varphi\pm\sqrt{\Delta}}{2}, \quad \Delta=\varphi^2-4(1-b^2)>0.
\end{equation}
Using (4.57) we also have $c$ in terms of $b$ as in (4.12). After substituting (4.58) into (4.55), one would obtain
\begin{equation}\label{4.59}
[\pm(\mu^2-1)\varphi\pm4\mu b+(\mu^2+1)\sqrt{\Delta}]b'-2(\mu^2 +1)b\sqrt{\Delta}\mp2\beta \varphi e^{2x}=0.
\end{equation}
We claim that the coefficient of $b'$ in (4.59) does not vanish in a non-empty open set. In fact, if it vanishes, then
\begin{equation*}
\begin{split}
&\pm(\mu^2-1)\varphi\pm4\mu b+(\mu^2+1)\sqrt{\Delta}=0,\\
&-2(\mu^2 +1)b\sqrt{\Delta}\mp2\beta \varphi e^{2x}=0.
\end{split}
\end{equation*}
Combining the above two equations, we conclude $\varphi^2+4b^2=0$. However, $\varphi^2+4b^2=0$ if and only if $\varphi=b=0$ which implies by (4.57) that $a=c$. This contradicts the Gauss equation (4.2). Thus, we can denote $b'=g(x,b)$, where $g$ is a differentiable function defined, from (4.59), by
\begin{equation*}
g(x,b)=\frac{2(\mu^2 +1)b\sqrt{\Delta}\mp2\beta \varphi e^{2x}}{\pm(\mu^2-1)\varphi\pm4\mu b+(\mu^2+1)\sqrt{\Delta}}.
\end{equation*}

Let $x_0$ be an arbitrary fixed point and consider the following initial value problem
\begin{equation}\label{4.60}
\begin{cases}
b'=g(x,b),\\
 b(x_0)=b_0.
 \end{cases}
\end{equation}
Since $b$ is a smooth function, then $g(x,b)$ and $\partial_b g(x,b)$ are continuous in some open rectangle
\begin{equation*}
R=\{(x,b):x_1<x<x_2,b_1<b<b_2\}
\end{equation*}
that contains the point $(x_0,b_0)$. Thus, by the fundamental existence and uniqueness theorem for ordinary differential equation, (4.60) has a unique solution in some closed interval $I=[b_0-\epsilon,b_0+\epsilon]$, where $\epsilon$ is a positive number. Moreover, $x_1$ and $x_2$ have to be chosen so that the strip $x_1<x<x_2$ intersects the domain of the solution of the generalized Camassa-Holm equation (1.3). Observe that substituting $\varphi$ into (4.59) produces (4.13).

The converse follows from a straightforward computation. This completes the proof.
\end{Proof}

\begin{Remark}
From (4.5) and (4.12) we obtain
\begin{equation}\label{4.61}
  H=\pm \frac{\sqrt{\Delta}}{2},
\end{equation}
thus $H$ is either positive or negative, which means that the sign of the mean value curvature is invariant on the domain of a given generic solution of the generalized Camassa-Holm equation (1.3).
\end{Remark}

\begin{Remark}
$\beta$ can be vanished. In fact, if $\beta=0$, it follows from (4.13) that
\begin{equation*}
  b\mu H=0,
\end{equation*}
which is a contradiction!
\end{Remark}

\section{Data availability}
No data was used for the research described in the article.

\end{document}